\documentclass{jltp}

\usepackage{graphics} 

\title{Annealing Effect for Supersolid Fraction in  $^4$He}

\author{Andrey Penzev, Yoshinori Yasuta, and Minoru Kubota}

\address{Institute for Solid State Physics, Tokyo University, \\
 Kashiwanoha 5-1-5, Kashiwa, Chiba 277-8581, Japan}

\runninghead{A.~Penzev, Y.~Yasuta, and M.~Kubota }{Annealing
Effect for Supersolid Signal in $^4$He}

\begin{document}

\maketitle

\begin{abstract}
 We report on experimental confirmation of the
non-classical rotational inertia (NCRI) in solid helium samples
originally reported by Kim and Chan. The onset of NCRI was
observed at temperatures below $\approx 400$ mK. The ac  velocity
for initiation of the NCRI suppression is estimated to be
$\approx$~10~$\mu$m/sec. After an additional annealing of the
sample at $T= 1.8$ K for 12 hours, $\sim$ 10\% relative increase
of NCRI fraction was observed. Then after repeated annealing with
the same conditions, the NCRI fraction was saturated. It differs
from Reppy's observation on a low pressure solid sample.

\end{abstract}

\section{INTRODUCTION}
 Recently the observation of NCRI\cite{legget} in solid $^4$He reported
 by Kim and Chan\cite{chan} for samples confined in Vyvor, in porous gold, and
 also in bulk solid.  The existence of NCRI was reported by Shirahama {\it et al.}\cite{shirahama}
 and by Rittner and Reppy\cite{reppy} with torsional oscillator (TO) measurements.
 We report on our independent experiment\cite{Penzev/Kubota} to check the existence of NCRI and related properties.
 In spite of very large differences of the surface to volume ratios in different samples,
 the NCRI fraction in low temperature limit was on the order of 1\% for
 all reported cases. This indicates the observed phenomenon is not for a
  surface related one, but for a uniform system over length scales,  6 nm and larger.   Nevertheless, a number of theoretical
 papers\cite{onsager,ceperley,svistunov,burovski,boninsegni,clark} suggest the superfluidity is unlikely to occur in a perfect crystal.

 A possibility for Bose Einstein condensation (BEC) of zero point vacancies was
  suggested by Andreev and Lifshitz\cite{andreev}. If observed decoupling is a simple consequence of BEC of zero point
 vacancies, then the supersolid fraction should monotonically decrease
 as the pressure of solid sample increases from the melting curve and
 should not strongly depend  on the sample quality. However the measured\cite{chan4}
 pressure dependence of NCRI fraction
 was nonmonotonous  with a maximum at $\approx 55$ bar.
 The absence of significant influence of annealing to the solid samples was pointed
out by Kim and Chan\cite{chan4}. 
  Bose condensation of other types of imperfections, for example,  interstitial-vacancy
  pair excitations as elementary excitations have been considered by Galli and
  Reatto\cite{Galli/Reatto}, and imperfections  at crystalline boundaries and
   dislocations are studied\cite{svistunov,burovski}. In the latter case we can expect some influence of sample quality.
 A recent experimental result supports such an idea that supersolid is
  preferred in  a disordered sample and annealing out such disorder would destroy supersolid sate.
 The complete disappearance  of NCRI  after annealing was observed by Rittner and
 Reppy\cite{reppy}.

 In the present report we confirm independently
 the existence of the NCRI  with TO
  technique  and also show a preliminary  result of the annealing influence to the
   bulk solid $^4$He samples, which indicate the opposite result to the low
   pressure  sample case reported by Ritter and Reppy\cite{reppy}.

\section{EXPERIMENTAL DETAILS}
We use a TO made of BeCu25 alloy. The sample bob has a replaceable
brass cover screwed and soldered with Wood's alloy forming a
cylindrical sample cell.
 Two types of cells have been used up to now. The difference between them is a small
capacitance pressure gauge similar to described in
Ref.~\onlinecite{gauge} mounted  only on the 1st cell. The
torsional rod is 2.5 mm outer diameter, 0.8 mm inner diameter and
20 mm long. The TO was mounted on a  copper block mechanical
isolator which was connected  to mixing chamber and was operated
at small ac velocities in constant amplitude mode. The 1st and 2nd
empty cell have resonant frequencies 1499.6 Hz and 1534.7 Hz,
respectively. The filling line was connected to the torsion rod
with minimum dead volumes to avoid the influence  of the solid
helium inside of the torsion rod. The solid sample was 8 mm in
diameter and 4 mm in height. Taking into account the volume of the
pressure gauge, the sample volume  was 160 mm$^3$ for the 1st cell
and 200 mm$^3$ for the 2nd.


 The solid samples were prepared  from commercially available
$^4$He ($\sim 0.3$ ppm $^3$He impurity) by blocked capillary
method at slow cooling (10-15 mK/min) along the melting curve to
avoid appearance of the pressure gradient at crystallization. The
sample 1 was annealed at $T=1.75$ K overnight before  the
measurements. The frequency change  due to the increase of inertia
moment by the sample is in reasonable agreement with molar volume
estimation (see Fig.~1).
\begin{figure}
\centerline{\scalebox{0.84}{\includegraphics*[312,225]{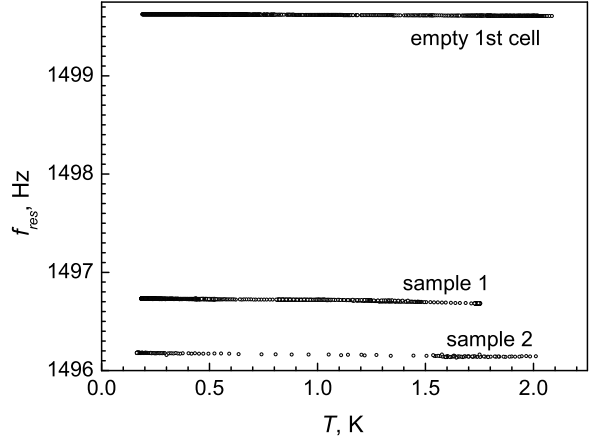}}}
 \caption{ The resonant frequency shifts due to
load of solid samples (1 and 2) as a function of temperature.}
\end{figure}

\begin{table}[h]
\centering
    \caption{Sample List}\label{tabl1}
    \vspace{0.1cm}
\begin{tabular*}{5 in}{c|c|c|c|c|c|c}

   \hline \hline

   \parbox{1.5 cm}{\vspace{0.2 cm}sample number \vspace{0.2cm}}&
   cell & $f_{res}$, Hz
   & \parbox{2cm}{initial pressure, bar} &  $T_m$, K
    & \parbox{2cm}{final pressure,
   bar}& $\Delta f$, Hz
   \\ \hline
  1 & 1st & 1499.6 & 71.7 & 2.08 & 40 & 2.96\\
  2 & 1st & 1499.6 & 90.5 & $-$ & $\approx$ 60 & 3.43\\
  3 & 2nd & 1534.7 & 74 & 2.15 & 43 & 3.76\\

  \hline \hline
\end{tabular*}
\end{table}
In the sample 1 measurement we utilized our pressure gauge, but
because of a large leak, appeared suddenly to the space between
gauge electrodes, it started to work as a density gauge with a
small sensitivity. We could still observe clear indication of the
crystallization finish of the sample and estimate the final
pressure of the sample. We could also see that the temperature
inside of the cell has the synchronous behavior for empty and
filled condition as the reading of a Matsushita carbon resistance
thermometer mounted on copper isolator. This thermometer was
calibrated against $^3$He melting curve thermometer below 1~K and
against calibrated Ge thermometer up to 4.99 K and then checked
with SRM768 fixed point device.

The measurements were performed at slow heating sweeps ($\sim 1$
mK/min) to take a lot of points for statistics because of problems
with amplitude stability caused by limited isolation of the
vibration. At the end of NCRI measurements series the sample was
heated to the melting point slowly and  temperature $T_m$, where
melting process sets in, was determined. Actually $T_m$ was
indicated by a sharp amplitude drop. Such a change is expected
 due to  dissipation by the nucleation fluctuations
 of the liquid.   A hysteresis
was observed between cooling and heating which is typical for the
first order transition. Finally the value of $T_m$ was determined
with accuracy $\pm 0.01$ K. The results for the different samples
are summarized in table \ref{tabl1}.

\section{RESULTS}

\begin{figure}[h]
\centerline{\scalebox{0.83}{\includegraphics*[312,
225]{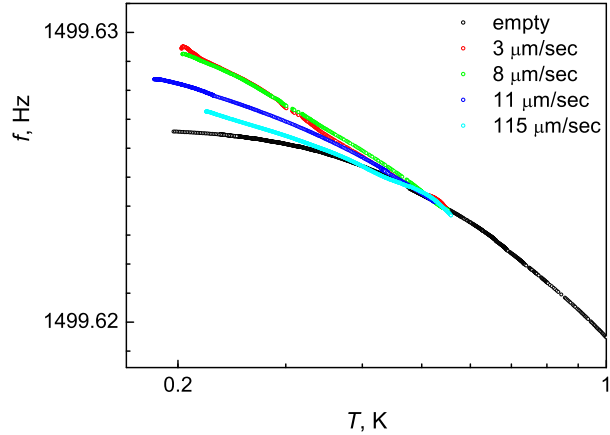}}} \caption{The temperature dependence of the
resonant frequency shifted to empty cell data for the sample 1 at
various ac velocities as indicated on the figure without
additional annealing.} \label{ncri1}
\end{figure}

\begin{figure}
\centerline{\scalebox{0.85}{\includegraphics*[280,
228]{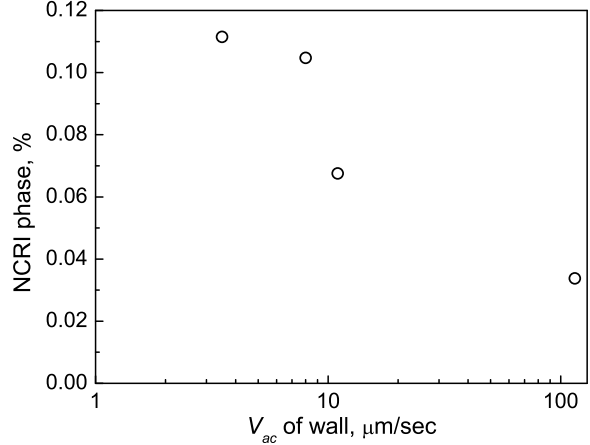}}} \caption{The NCRI fraction {\it vs.} $V_{ac}$
for the sample 1 at $T=200$ mK. The data for 11 $\mu$m/sec were
extrapolated to 200 mK.} \label{critical}
\end{figure}

We show the results of the NCRI measurements. Fig. \ref{ncri1}
shows temperature dependence of the NCRI, resonant frequency at
various ac velocities of the cell wall for  sample 1 without
additional annealing.  Below 400 mK clear deviation from empty
cell data observed. This indicates sample moment of inertia
decoupling and we consider it as NCRI.  NCRI fraction for the
sample 1 was $\approx 0.1$ \% at 200 mK, that is rather large in
comparison with other experimental
data\cite{chan,shirahama,reppy,chan4}. Fig.~\ref{critical} shows
dependence of NCRI fraction at 200 mK on ac wall velocity  and its
suppression  above $\sim 10 \mu$m/sec.

Observed temperature dependence of NCRI fraction has qualitatively
the same shape for all the samples.
 The sample 2 was measured only  once  at $V_{ac}=8 ~\mu$m/sec
  and also exhibited
  high onset temperature $\approx 350$ mK and NCRI fraction
  $\approx 0.14$ \% at $T=130$ mK with extrapolation of the empty cell resonant frequency down to
lower temperatures.

\begin{figure}[h]
\centerline{\includegraphics*[312, 225]{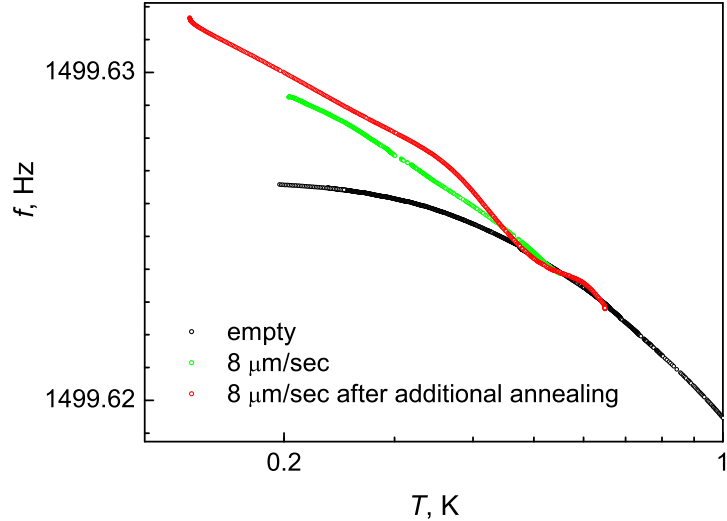}} \caption{The
difference between NCRI before(green) and after(red) first
additional anealling.} \label{anneal}
\end{figure}
For the sample 3 which was a bulk cylinder, the effect starts
below 250 mK and maximum NCRI fraction was only 0.08\% at 75 mK.
Such reduction of NCRI for bulk cylinder is in agreement with data
by Shirahama {\it et al.}\cite{shirahama}.

The additional annealing was performed for the sample 1 at $T=1.8$
K for $\approx 12$ hours. After annealing we performed
measurements at 8 $\mu$m/sec. We observed relative increase of
NCRI fraction up to 10 \% (see Fig.~\ref{anneal}). After the
second and third additional annealing no observable change
appeared.

\section{SUMMARY}

The existence  of NCRI was confirmed  by TO technique. The
measured NCRI fraction is in reasonable agreement with Kim and
Chan data\cite{chan4}. The critical velocity is found to be on the
order of 10 $\mu$m/sec.  Annealing
   near the melting curve resulted in increase of NCRI fraction and further
    annealing made an saturation in it, but NCRI fraction
     never decreased by annealing as in low pressure samples.

\section*{ACKNOWLEDGMENTS}
 Authors acknowledge technical support by T. Igarashi.
 They enjoy discussions with R. Mueller and Y.~Karaki and their help in fixed point device
 measurements and resistance thermometer calibration.
 This research was partially
supported by JSPS. A.P. would like to thank ISSP, U-Tokyo for
hospitality during his stay.


\begin{thebibliography}{9}

\bibitem{legget} A. Leggett, {\it Phys. Rev. Lett.} {\bf 25}, 1543
(1970).

\bibitem{chan}E. Kim and M.H.W. Chan, {\it Nature} {\bf 427}, 225
(2004); {\it J. Low Temp. Phys.} {\bf 138}, 859 (2005);
 {\it Science} {\bf 305}, 1941 (2004);
Science Express 1101501.

\bibitem{shirahama} K. Shirahama {\it et al.}, {\it Bull. Am. Phys. Soc.} {\bf 51},
450 (2006) and private comm.

\bibitem{reppy} A.S.C. Rittner and J.D. Reppy, cond-mat/0604528.

\bibitem{Penzev/Kubota} A. Penzyev and M. Kubota, http://online.itp.ucsb.edu/online/smatter\_m06/kubota/.

\bibitem{onsager} O. Pernose and L. Onsager, {\it Phys. Rev.} {\bf
104}, 576 (1956).

\bibitem{ceperley} D.M. Ceperley and B. Bernu, {\it Phys. Rev.
Lett.} {\bf 93}, 155303 (2004).

\bibitem{svistunov} N. Prokof'ev and B. Svistunov, {\it Phys. Rev.
Lett.} {\bf 94}, 155302 (2005).

\bibitem{burovski} E. Burovski {\it et al.}, {\it Phys. Rev. Lett.} {\bf
94}, 165301 (2005).

\bibitem{boninsegni} M. Boninsegni {\it et al.},
{\it Phys. Rev. Lett.} {\bf 96}, 105301 (2006).

\bibitem{clark}  B.K. Clark and D.M. Ceperley, {\it Phys. Rev.
Lett.} {\bf 96}, 105302 (2006).



\bibitem{andreev} A.F. Andreev and I.M. Lifshitz, {\it Sov.
Phys. JETP} {\bf 29}, 1107 (1969).

\bibitem{chan4} E. Kim and M.H.W. Chan, cond-mat/0605680.

\bibitem{Galli/Reatto} D.E. Galli and L. Reatto, {\it Phys. Rev. Lett.} {\bf
96}, 165301 (2006).

\bibitem{gauge} Y. Morii and E.D. Adams, {\it Rev. Sci. Instrum.} {\bf 52}, 1779 (1981).




\end{thebibliography}
\end{document}